\documentclass[copyright,creativecommons]{eptcs}

\providecommand{\event}{ACL2 2014} % Name of the event you are submitting to
\usepackage{breakurl}             % Not needed if you use pdflatex only.

\usepackage[fleqn]{amsmath}
\usepackage{amsthm}
\usepackage{amssymb}
\newtheorem{theorem}{Theorem}

\usepackage{graphicx}

\newcommand{\ceil}[1]{\lceil #1 \rceil}

\usepackage{listings}
\usepackage{txfonts}

\usepackage{url}
\newcommand{\keywordname}{Keywords:}
\newcommand{\keywords}[1]{\par\addvspace\baselineskip\noindent\keywordname\enspace\ignorespaces#1}

\begin{document}

\title{Equivalence of the Traditional and Non-Standard Definitions of Concepts from Real Analysis}
\def\titlerunning{Traditional and Non-Standard Definitions in Real Analysis}
\def\authorrunning{John Cowles \& Ruben Gamboa}

\author{John Cowles
\institute{University of Wyoming \\
Laramie, WY, USA}
\email{cowles@uwyo.edu}
\and
Ruben Gamboa
\institute{University of Wyoming \\
Laramie, WY, USA}
\email{ruben@uwyo.edu}
}

\maketitle

\begin{abstract}
ACL2(r) is a variant of ACL2 that supports the irrational real and complex numbers. Its logical foundation is based on internal set theory (IST), an axiomatic formalization of non-standard analysis (NSA). Familiar ideas from analysis, such as continuity, differentiability, and integrability, are defined quite differently in NSA---some would argue the NSA definitions are more intuitive. In previous work, we have adopted the NSA definitions in ACL2(r), and simply taken as granted that these are equivalent to the traditional analysis notions, e.g., to the familiar $\epsilon$-$\delta$ definitions. However, we argue in this paper that there are circumstances when the more traditional definitions are advantageous in the setting of ACL2(r), precisely because the traditional notions are classical, so they are unencumbered by IST limitations on inference rules such as induction or the use of pseudo-lambda terms in functional instantiation. To address this concern, we describe a formal proof in ACL2(r) of the equivalence of the traditional and non-standards definitions of these notions.

\end{abstract}

\section{Introduction}
\label{intro}

ACL2(r) is a variant of ACL2 that has support for reasoning about the
irrational numbers.  The logical basis for ACL2(r) is \emph{non-standard
analysis} (NSA), and in particular, the axiomatic treatment of NSA
developed as \emph{internal set theory} (IST)~\cite{Nel:nsa}.  Traditional notions
from analysis, such as limits, continuity, and derivatives, have
counterparts in NSA

Previous formalizations of NSA typically prove that these definitions
are equivalent early on.  We resisted this in the development of
ACL2(r), preferring simply to state that the NSA notions were the
``official'' notions in ACL2(r), and that the equivalence to the usual
notions was a ``well-known fact'' outside the purview of ACL2(r).
In this paper, we retract that statement for three reasons. 

First,
the traditional notions from real analysis require the use of quantifiers.  For
instance, we say that a function $f$ has limit $L$ as $x$ approaches
$a$ iff 
\begin{equation*}
\forall\epsilon>0, \exists \delta>0 \text{ such that }
|x-a|<\delta \Rightarrow |f(x)-L|<\epsilon.
\end{equation*}
While ACL2(r) has only limited support for quantifiers, this support
is, in fact, sufficient to carry out the equivalence proofs. However,
it should be noted that the support depends on recent enhancements to
ACL2 that allow the introduction of Skolem functions with
non-classical bodies. So, in fact, it is ACL2's improved but still
modest support for quantifiers that is sufficient.  That story
is interesting in and of itself.

Second, the benefit of formalization in general applies to this case,
as the following anecdote illustrates.  While trying to update the
proof of the Fundamental Theorem of Calculus, we were struggling to
formalize the notion of \emph{continuously differentiable}, i.e., that
$f$ is differentiable and $f'$ is continuous.  To talk about the class
of differentiable functions in ACL2(r), we use an \texttt{encapsulate}
event to introduce an arbitrary differentiable functions. It would be very
convenient to use the existing \texttt{encapsulate} for differentiable
functions, and prove as a theorem that the derivative was
continuous. That is to say, it would be very convenient if all
derivative functions were continuous. Note: we mean ``derivative''
functions, not ``differentiable'' functions. The latter statement had
previously been proved in ACL2(r).

Encouraged by Theorem 5.6 of~\cite{Nel:nsa}, one of us set out to
prove that, indeed, all derivatives of functions are continuous.
\begin{theorem}[5.6, \cite{Nel:nsa}]
Let $f:I\rightarrow\mathbb{R}$ where $I$ is an interval. If $f$ is
differentiable on $I$, then $f'$ is continuous on $I$.
\end{theorem}
Nelson's proof of this theorem begins with the following statement:
\begin{quote}
  We know that
\begin{equation}
\label{eqn-deriv}
\forall^\text{st}x \forall x_1 \forall x_2 \left\{ x_1 \approx x
  \wedge x_2 \approx x \wedge x_1 \ne x_2 \Rightarrow
  \frac{f(x_2)-f(x_1)}{x_2-x_1} \approx f'(x)\right\}.
\end{equation}
\end{quote}
This is, in fact, plausible from the definition of continuity, which
is similar but with $x$ taking the place of $x_2$.  The remainder of
the proof was ``trivially'' (using the mathematician's sense of the
word) carried out in ACL2(r), so only the proof of this known fact
remained. The hand proof for this fact was tortuous, but eminently
plausible. Unfortunately, the last step in the proof failed, because
it required that $y\cdot y_1 \approx y\cdot y_2$ whenever $y_1
\approx y_2$---but this is true only when $y$ is known to be limited.

The other of us was not fooled by Theorem 5.6: What about the function
$x^2 \sin(1/x)$? The discrepancy was soon resolved. Nelson's
definition of derivative in~\cite{Nel:nsa} is precisely
Equation~\ref{eqn-deriv}. No wonder this was a known fact! And the
problem is that Equation~\ref{eqn-deriv} is equivalent to the notion
of continuously differentiable, and \emph{not equivalent} to
the usual notion of differentiability. But in that case, how are we to
know if theorems in ACL2(r) correspond to the ``usual'' theorems in
analysis.  I.e., what if we had chosen Equation~\ref{eqn-deriv} as the
definition of derivative in ACL2(r)? Preventing this situation from
reoccurring is the second motivator for proving the equivalence of the
definitions in ACL2(r) once and for all.

Third, the NSA definitions are non-classical; i.e., they use notions
such as ``infinitely close'' and ``standard.''  Indeed, it is these
non-classical properties that make NSA such a good fit for the
equational reasoning of ACL2(r). However, non-classical functions are
severely limited in ACL2(r): Induction can be used to prove theorems 
using non-classical functions only up to standard values of the free
variables, and function symbols may not map to pseudo-lambda
expressions in a functional instantiation~\cite{GC:acl2r-theory}. 
As a practical consequence
of these restrictions, it is impossible to prove that
$\frac{d(x^n)}{dx} = n \cdot x^{n-1}$ by using the product rule and
induction in ACL2(r). In~\cite{Gam:dissertation}, for example, this is
shown only for standard values of $n$. However, using the traditional
notion of differentiability, the result does follow from induction.
This, too, would have been reason enough to undertake this work.

It should be emphasized that the main contribution of this paper is the
formalization in ACL2(r) of the results described in this paper.  The
actual mathematical results are already well-known in the non-standard
analysis community.  Moreover, some of these equivalence results were
formalized mechanically as early as~\cite{BaBl:nsa}.  The novelty here
is the formalization in ACL2(r), which complicates things somewhat 
because of the poor support for (even first-order) set theory.

The rest of this paper is organized as follows.  In
Section~\ref{series}, we discuss equivalent definitions regarding
convergence of series\footnote{Readers who attended the ACL2
  Workshop in 2013 will recognize many of the results in this section,
because they were presented in a Rump Session
there.}. Section~\ref{limits} considers the limit of a function at a
point. The results in this section are used in
Section~\ref{continuity} to show that the notions of continuity at a 
point are also equivalent. This leads into the discussion of
differentiability in Section~\ref{differentiability}. Finally,
Section~\ref{integrability} deals with the equivalent definitions of
Riemann integration.

\section{Convergence of Series}
\label{series}

In this section, we show that several definitions of convergence are in
fact equivalent.  In particular, we will consider the traditional
definitions, e.g., as found in~\cite{Rudin:analysis}, and
the corresponding concepts using non-standard analysis,
e.g., as found in~\cite{Robinson:nsa}.

We start with the constrained function \texttt{Ser1},
which represents an arbitrary sequence; i.e., it is a fixed but arbitrary
function that maps the natural numbers to the reals.  Moreover,
\texttt{Ser1} is assumed to be a classical function---otherwise, some of
the equivalences do not hold.
Similarly, the function \texttt{sumSer1-upto-n} defines the partial sum of \texttt{Ser1},
i.e., the sum of the values of \texttt{Ser1} from $0$ to \texttt{n}.

The first definition of convergence is the traditional one due to 
Weierstrass: 
\begin{equation*}
(\exists L) (\forall \epsilon) (\exists M) (\forall n) 
(n > M \Rightarrow | \sum_{i=0}^{n}{a_i} - L| < \epsilon).  
\end{equation*}
In ACL2, we can write the innermost quantified subformula of this definition as 
follows:
\begin{lstlisting}
  (defun-sk All-n-abs-sumSer1-upto-n-L<eps (L eps M)
    (forall n (implies (and (standardp n)
                            (integerp n)
                            (> n M))
                       (< (abs (- (sumSer1-upto-n n) L))
                          eps))))
\end{lstlisting}
This version of the definition restricts \texttt{n} to be a standard
integer, which makes it a non-classical formula.  A different version omits this 
requirement, and it is a more direct translation of Weierstrass's criterion.
\goodbreak
\begin{lstlisting}
  (defun-sk Classical-All-n-abs-sumSer1-upto-n-L<eps (L eps M)
    (forall n (implies (and (integerp n)
                            (> n M))
                       (< (abs (- (sumSer1-upto-n n) L))
                          eps))))
\end{lstlisting}
ACL2 can verify that these two conditions are equal to each other,
but only when the parameters \texttt{L}, \texttt{eps}, and \texttt{M}
are standard.  This follows because \texttt{defchoose} is guaranteed to
choose a standard witness for classical formulas and standard parameters.
More precisely, the witness function is a classical formula is also classical,
all all classical functions return standard values for standard inputs~\cite{GC:acl2r-theory}.
Once this basic equivalence is proved, it follows
that both the classical and non-classical versions of 
Weierstrass's criterion are equivalent.  It is only necessary to
add each of the remaining quantifiers one by one.

We note in passing that the two versions of Weierstrass's criterion 
are \emph{not} equivalent for non-standard values of the parameters
\texttt{L}, \texttt{eps}, and \texttt{M}.  
Consider, for example,
the case when \texttt{eps} is infinitesimally small.  It is straightforward
to define the sequence $\{a_n\}$ such that the partial sums 
are given by $\sum_{i=1}^{n}{a_i} = 1/n$, clearly converging
to $0$.  Indeed, for any $\epsilon>0$ there is an $N$ such that
for all $m > N$, $\sum_{i=1}^{m}{a_i} = 1/m < 1/N < \epsilon$.
However, for infinitesimally small $\epsilon$, the resulting $N$
is infinitesimally large.  This is fine using the second (classical)
version of Weierstrass's criterion, but not according to the first,
since for all standard $N$, $1/N > \epsilon$, so no standard $N$ can
satisfy the criterion.  However, the two criteria are equivalent when
written as sentences, i.e., when they have no free variables.

Note that the only difference between the two versions of 
Weierstrass's criterion is that one of them features only standard
variables, whereas the other features arbitrary values for all
quantified variables.  Using the shorthand $\forall^\text{st}$ and
$\exists^\text{st}$ to introduce quantifiers for standard variables,
the two versions of the criteria can be written as follows:
\begin{itemize}
\item $(\exists^\text{st} L) (\forall^\text{st} \epsilon) (\exists^\text{st} M) (\forall^\text{st} n) 
(n > M \Rightarrow | \sum_{i=0}^{n}{a_i} - L| < \epsilon)$
\item $(\exists L) (\forall \epsilon) (\exists M) (\forall n) 
(n > M \Rightarrow | \sum_{i=0}^{n}{a_i} - L| < \epsilon)$  
\end{itemize}
It is obvious that these two statements are extreme variants, and
that there are other possibilities mixing the two types of quantifiers.
Indeed, we verified with ACL2 that the following versions are also
equivalent to the above:
\begin{itemize}
\item $(\exists^\text{st} L) (\forall^\text{st} \epsilon) (\exists^\text{st} M) (\forall n) 
(n > M \Rightarrow | \sum_{i=0}^{n}{a_i} - L| < \epsilon)$
\item $(\exists^\text{st} L) (\forall^\text{st} \epsilon) (\exists M) (\forall n) 
(n > M \Rightarrow | \sum_{i=0}^{n}{a_i} - L| < \epsilon)$
\end{itemize}

The last two versions of Weierstrass's criterion are useful, because
they are easier to show equivalent to the typical non-standard criterion 
for convergence:
$(\exists L) (\forall n) (large(n) \Rightarrow \sum_{i=0}^{n}{a_i}  \approx L)$,
i.e., for large values of $n$, $\sum_{i=0}^{n}{a_i}$ is infinitely close to $L$.  
This is the convergence criterion used
in~\cite{Gam:dissertation}, for example, where power series are used
to introduce functions such as $e^x$.

There is another statement of the non-standard convergence criterion
that appears weaker: 
\begin{equation*}
(\exists L) (\exists M) (large(M) \wedge (\forall n) (n > M
\Rightarrow \sum_{i=0}^{n}{a_i} \approx L)).
\end{equation*}
This version does not require that $\sum_{i=0}^{n}{a_i} $ is close to $L$ for all large $n$, only
that this is true for $n$ larger than some large $M$.
We have shown in ACL2 that these statements
are in fact equivalent to Weierstrass's criterion for convergence.
In fact, since $\{a_n\}$ is a classical sequence, the value of $L$ is guaranteed
to be standard, so we can replace $(\exists L)$ with $(\exists^\text{st} L)$
in both of the non-classical convergence criteria given above and 
still retain equivalence.

When the sequence is composed of non-negative numbers, we can make
even stronger guarantees.  Let $\{b_n\}$ be such a sequence, which
we introduce into ACL2 as the constrained function \texttt{Ser1a}.
All the previous results about \texttt{Ser1}---i.e., about $\{a_n\}$---apply
to \texttt{Ser1a}, and we can carry over these proofs in ACL2 by
using functional instantiation.

Using the non-standard criterion for convergence, we can easily see that
if $\sum_{i=0}^{\infty}b_n$ converges, then $\sum_{i=0}^{N}{b_i}$ is not infinitely large, where $N$ is a fixed
but arbitrary large integer\footnote{The ACL2 constant \texttt{(i-large-integer)} 
is often used to denote an otherwise unspecified large integer, and that is what we use
in this case.}. This simply follows from the facts that $\sum_{i=0}^{N}{b_i}\approx L$ and
$L$ is standard.

The converse of this fact is also true: if $\sum_{i=0}^{N}{b_i}$ is not infinitely large, then
$\sum_{i=0}^{\infty}b_n$ converges.  This is harder to prove formally.  The key idea is 
as follows. Since $\sum_{i=0}^{N}{b_i}$ is not infinitely large, then 
$\sum_{i=0}^{N}{b_i}$ must be close to
an unique standard real number, i.e., $\sum_{i=0}^{N}{b_i} \approx L$ for some 
standard $L$.  $\sum b_i$ is monotonic, so for any standard $n$, 
$\sum_{i=0}^{n}{b_i} \le \sum_{i=0}^{N}{b_i}$.  And since $L$ is the
unique real number that is close to $\sum_{i=0}^{N}{b_i}$, we can
conclude that $\sum_{i=0}^{n}{b_i} \le L$ for all standard $n$.
Using the non-standard transfer principle, this is sufficient to conclude
that $\sum_{i=0}^{n}{b_i} \le L$ for all $n$, not just the standard ones.
Using monotonicity once more, it follows that 
whenever $n>N$, $\sum_{i=0}^{n}{b_i} \approx L$, which is precisely
the (weak) non-standard convergence criterion above.  Thus,
the series  $\sum_{i=0}^{N}{b_i}$ converges, according to any of the criteria above.

Similar results hold for divergence to positive infinity.  Let $\{c_n\}$ be an arbitrary sequence.
Weierstrass's criterion is given by 
$(\forall^\text{st} B) (\exists^\text{st} M) (\forall^\text{st} n) (n > M \Rightarrow \sum_{i=0}^{n}{c_i} > B)$.
As before, for classical $\{c_n\}$ this is equivalent to a criterion with quantifiers over
all reals, not just the standard ones: 
$(\forall B) (\exists M) (\forall n) (n > M \Rightarrow \sum_{i=0}^{n}{c_i} > B)$.
And just as before, other variants (with $B$ and $M$ standard or just $B$ standard) are also 
equivalent.  Moreover, these are equivalent to the non-standard criterion for divergence to positive infinity,
namely that $(\forall n) (large(n) \Rightarrow
large(\sum_{i=0}^{n}{c_i}))$. A seemingly weaker version of this
criterion is also equivalent, where it is only necessary that $c_n$ is large for all $n$ beyond
a given large integer:  $(\exists M) (large(M) \wedge (\forall n) (n > M \Rightarrow large(\sum_{i=0}^{n}{c_i})))$.
Finally, if the sequence $\{c_n\}$ consists of non-negative reals, then it is even easier
to show divergence.  It is only necessary to test whether $large(\sum_{i=0}^{N}{c_i})$ where $N$ 
is an arbitrary large integer, and as before we choose the ACL2 constant \texttt{i-large-integer}
for this purpose.

\section{Limits of Functions}
\label{limits}

In this section, we consider the notion of limits. In particular, we
show that the following three notions are equivalent (for standard
functions and parameters):
\begin{itemize}
\item The non-standard definition (for standard parameters $a$ and $L$): 
\begin{equation*}
\lim_{x \rightarrow a} f(x) = L \Leftrightarrow  
\left((\forall x) (x \approx a \wedge x \ne a \Rightarrow f(x) \approx L\right)).
\end{equation*}

\item The traditional definition over the classical reals: 
\begin{equation*}
\lim_{x \rightarrow a} f(x) = L \Leftrightarrow  
\left((\forall^\text{st} \epsilon > 0) (\exists^\text{st} \delta>0)  (0<|x-a|<\delta
  \Rightarrow | f(x) - L| < \epsilon)\right).
\end{equation*}

\item The traditional definition over the hyperreals: 
\begin{equation*}
\lim_{x \rightarrow a} f(x) = L \Leftrightarrow  
\left((\forall \epsilon > 0) (\exists \delta>0)  (0<|x-a|<\delta
  \Rightarrow | f(x) - L| < \epsilon)\right).
\end{equation*}
\end{itemize}

We begin by assuming the non-standard definition, which can be
introduced in ACL2(r) by encapsulating the function $f$, its domain,
and the limit function $L$, so that $\lim_{x \rightarrow a} f(x) =
L(a)$.  The first step is to observe that $a\approx b$ is a shorthand
notation for the condition that $|a-b|$ is infinitesimally small.
Moreover,  if $\epsilon>0$ is standard, then it must be (by
definition) larger than any infinitesimally small number.  Thus, we
can prove that
\begin{equation*}
(\forall^\text{st} \epsilon > 0) \left((\forall x) (x \approx a \wedge x \ne a
  \Rightarrow |f(x) - L(a)| < \epsilon\right)).
\end{equation*}
Similarly, if $\delta>0$ is infinitesimally small, then $|x - a| <
\delta$ implies that $x \approx a$.  It follows then that
\begin{equation*}
(\forall^\text{st} \epsilon > 0) 
(\forall \delta>0)
  \left(small(\delta) \Rightarrow(\forall x)  \left(0< |x - a| < \delta \wedge x \ne a
  \Rightarrow |f(x) - L(a)| < \epsilon\right)\right).
\end{equation*}
It is an axiom of ACL2(r) that there exists a positive infinitesimal,
namely \texttt{(/ (i-large-integer))}.  Consequently, we can
specialize the previous theorem with the constant $\delta_0$ (i.e., \texttt{(/ (i-large-integer))}).
\begin{equation*}
(\forall^\text{st} \epsilon > 0) 
  \left(0<\delta_0 \wedge small(\delta_0) \wedge (\forall x)  \left(0< |x - a| < \delta_0 \wedge x \ne a
  \Rightarrow |f(x) - L(a)| < \epsilon\right)\right).
\end{equation*}
Using ACL2 terminology, the specific number $\delta_0$ can be
generalized to yield the following theorem:
\begin{equation*}
(\forall^\text{st} \epsilon > 0) 
(\exists \delta>0)
\left(
  (\forall x)  \left(0< |x - a| < \delta \wedge x \ne a
  \Rightarrow |f(x) - L(a)| < \epsilon\right)\right).
\end{equation*}
Note that the statement inside the $\forall^\text{st}$ is classical;
i.e., it does not use any of the notions from NSA, such as standard,
infinitesimally close, infinitesimally small, etc.  Consequently, we
can use the transfer principle so that the quantifier ranges over all
reals instead of just the standard reals.  This results in the
traditional definition of limits over the hyperreals:
\begin{equation*}
(\forall \epsilon > 0) 
(\exists \delta>0)
\left(
  (\forall x)  \left(0< |x - a| < \delta \wedge x \ne a
  \Rightarrow |f(x) - L(a)| < \epsilon\right)\right).
\end{equation*}
The transfer can also be used in the other direction. 
The introduction of the existential quantifier is done via
\texttt{defun-sk}, and ACL2(r) introduces such quantifiers by creating a
Skolem choice function $\delta(a,\epsilon)$ using \texttt{defchoose}. 
Since the criteria used to define this Skolem function are classical,
\texttt{defchoose} introduces the Skolem function itself as
classical. That means that when $a$ and $\epsilon$ are standard, so is
$\delta(a, \epsilon)$.  This observation is sufficient to show that
$\lim_{x\rightarrow a} f(x) = L(a)$, using the traditional definition
over the classical reals:
\begin{equation*}
(\forall^\text{st} \epsilon > 0) 
(\exists^\text{st} \delta>0)
\left(
  (\forall x)  \left(0< |x - a| < \delta \wedge x \ne a
  \Rightarrow |f(x) - L(a)| < \epsilon\right)\right).
\end{equation*}

It is worth noting that this last theorem is not obviously weaker or
stronger than the previous one, where the quantifiers range over all
reals, not just the standard ones.  The reason is that the $\forall$
quantifier ranges over more values than $\forall^\text{st}$, so it
would appear that using $\forall$ instead of $\forall^\text{st}$
yields a stronger result. However, this advantage is lost when one
considers the $\exists$ quantifier, since $\exists^\text{st}$ gives an
apparently stronger guarantee.  In actual fact, the two statements are
equivalent, since the transfer principle can be used to guarantee that
the value guaranteed by $\exists$ can be safely assumed to be
standard.

To complete the proof, we need to show that if $\lim_{x\rightarrow a}
f(x) = L(a)$, using the traditional definition over the standard reals,
then $\lim_{x\rightarrow a} f(x) = L(a)$ using the non-standard
definition.  To do this, we introduce a new \texttt{encapsulate} where
$f$ is constrained to have a limit using the traditional definition
over the standard reals.  We then proceed as follows.  First, fix
$\epsilon$ so that it is positive and standard.  From the (standard
real) definition of limit, it follows that
\begin{equation*} (\exists^\text{st} \delta > 0) (\forall x) \left(0<|x-a|<\delta
  \Rightarrow | f(x) - L(a)| < \epsilon\right).
\end{equation*}
Now suppose that $\delta_0$ is a positive, infinitesimally small
number. It follows that $\delta_0 < \delta$ for any positive, standard
$\delta$. In particular, this means that 
\begin{equation*}0 < \delta_0 \wedge (\forall x) \left(0<|x-a|<\delta_0
  \Rightarrow | f(x) - L(a)| < \epsilon\right).
\end{equation*}
Since $\delta_0$ is an arbitrary positive infinitesimal, we can
generalize it as follows:
\begin{equation*}(\forall \delta > 0) \left(small(\delta) \Rightarrow (\forall x) \left(0<|x-a|<\delta
  \Rightarrow | f(x) - L(a)| < \epsilon\right)\right).
\end{equation*}
Next, we remove the universal quantifier on $x$. This step does not
have a dramatic impact on the mathematical statement, but it is more
dramatic in ACL2(r), since it opens up a function introduced with
\texttt{defun-sk}:
\begin{equation*}(\forall \delta > 0) \left(small(\delta) \Rightarrow \left(0<|x-a|<\delta
  \Rightarrow | f(x) - L(a)| < \epsilon\right)\right).
\end{equation*}
Recall that $x \approx a$ is a shorthand for $|x-a|$ is
infinitesimally small.  Thus, the theorem implies the following 
\begin{equation*}(\forall \delta > 0) \left(small(\delta) \Rightarrow \left(x \approx
    a \wedge x \ne a  \Rightarrow | f(x) - L(a)| <
    \epsilon\right)\right).
\end{equation*}
At this point, the variable $\delta$ is unnecessary, so we are
left with the following:
\begin{equation*} x \approx a \wedge x \ne a  \Rightarrow | f(x) - L(a)| <
\epsilon.
\end{equation*}
Now, recall that we fixed $\epsilon$ to be an arbitrary, positive,
standard real. This means that what we have shown is actually the
following:
\begin{equation*}(\forall^\text{st} \epsilon) \left(x \approx a \wedge x \ne a  \Rightarrow |f(x) - L(a)| <
\epsilon\right).
\end{equation*}
To complete the proof, it is only necessary to observe that if
$|x-y|<\epsilon$ for all standard $\epsilon$, then $x \approx y$.  We
prove this in ACL2(r) by finding an explicit standard $\epsilon_0$ such
that if $x \not\approx y$, then $|x-y| > \epsilon_0$.  The details of
that proof are tedious and not very elucidating, so we omit them from
this discussion\footnote{The interested reader can consult the definition of
\texttt{standard-lower-bound-of-diff} which produces the constant
$\epsilon_0$ mentioned above, and the lemmas \texttt{standards-are-in-order-2},
\texttt{standards-are-in-order}, \texttt{rlfn-classic-has-limits-step-3},
and the more trivial lemmas leading up to the main theorem \texttt{rlfn-classical-has-a-limit-using-nonstandard-criterion}.}.  
Once that lemma is proved, however, it follows that
$\lim_{x\rightarrow a} f(x) = L(a)$ using the non-standard definition:
\begin{equation*}x \approx a \wedge x \ne a  \Rightarrow f(x) \approx
  L(a).
\end{equation*}

These results show that the three definitions of limit are indeed
equivalent, at least when $f$ and $L$ are classical, and $a$ is
standard.

\section{Continuity of Functions}
\label{continuity}

Now we consider the notion of continuity. The function $f$ is said to
be continuous at $a$ if $\lim_{x \rightarrow a} f(x) = f(a)$.  Since
this uses the notion of limit, it is no surprise that there are three
different characterizations which are equivalent (for standard
functions and parameters):
\begin{itemize}
\item The non-standard definition (for standard parameter $a$): 
\begin{equation*}f \text{ is continuous at } a \Leftrightarrow  
\left((\forall x) x \approx a \wedge x \ne a \Rightarrow f(x) \approx
  f(a)\right).
\end{equation*}
\item The traditional definition over the classical reals: 
\begin{equation*}f \text{ is continuous at } a \Leftrightarrow  
\left((\forall^\text{st} \epsilon > 0) (\exists^\text{st} \delta>0)  (0<|x-a|<\delta
  \Rightarrow | f(x) - f(a)| < \epsilon)\right).
\end{equation*}
\item The traditional definition over the hyperreals: 
\begin{equation*}f \text{ is continuous at } a \Leftrightarrow  
\left((\forall \epsilon > 0) (\exists \delta>0)  (0<|x-a|<\delta
  \Rightarrow | f(x) - f(a)| < \epsilon)\right).
\end{equation*}
\end{itemize}

What this means is that the notion of continuity can be completely
reduced to the notion of limits.  In particular, the results from
Section~\ref{limits} can be functionally instantiated to derive the
results for continuity.  It is only necessary to instantiate both
functions $f(x)$ and $L(x)$ to the same function $f(x)$.

\section{Differentiability of Functions}
\label{differentiability}

Next, we consider differentiability. At first sight, it appears that
we can also define differentiability in terms of limits.  After all,
$f'$ is the derivative of $f$ iff
\begin{equation*}
  \lim_{\epsilon \rightarrow 0} \frac{f(x+\epsilon) - f(x)}{\epsilon} = f'(x).
\end{equation*}
The problem, however, is that the difference quotient on the left of the
equation is a function of both $x$ and $\epsilon$, and having free
variables complicates functional instantiation when non-classical
functions are under consideration.
So we chose to prove this result essentially from scratch, although
the pattern is very similar to the equivalence of limits.

Before proceeding, however, it is worth noting one other equivalence
of interest.  The non-standard definition of differentiability is as
follows: 
\begin{gather*}
standard(a) \wedge x_1 \approx a \wedge x_1 \ne a \wedge x_2 \approx a \wedge x_2 \ne a \Rightarrow \\
\qquad\left(\neg large\left(\frac{f(x_1) - f(a)}{x_1 - a}\right) \wedge
  \frac{f(x_1) - f(a)}{x_1 - a} \approx \frac{f(x_2) - f(a)}{x_2 - a}\right).
\end{gather*}
The form of this definition was chosen because it does not have a
dependency on $f'$, so it can be applied to functions even when their
derivative is unknown.  However, when $f'$ is known, a simpler
definition can be used:
\begin{equation*}
standard(a) \wedge x \approx a \wedge x \ne a \Rightarrow \left(\frac{f(x) - f(a)}{x - a} \approx f'(a)\right).
\end{equation*}
In fact, this latter form is the definition of differentiability that
was used in~\cite{ReGa:automatic-differentiator}.  In that context,
ACL2(r) was able to automatically define $f'$ from the definition of
$f$, so $f'$ was always known and the simpler definition was
appropriate.

So the first result we show is to relate the definitions of
differentiable and derivative.  To do so, we can begin
with a differentiable function $f$ and define $f'$ (for standard $a$)
as follows:
\begin{equation*}
f'(a) \equiv standard\text{ }part\left(\frac{f(a+\epsilon) - f(a)}{\epsilon}\right)
\end{equation*}
where $\epsilon$ is a fixed but arbitrary, positive, small real, e.g.,
\texttt{(/ (i-large-integer))}.
By assumption, the difference quotient at $a$ is not large for $x_1 = a +
\epsilon$.  Since $f'(a)$ is defined as the standard part of the
difference quotient, it follows that it really is close to the difference quotient,
so $f'$ really is the derivative of $f$.

Conversely, suppose $f'$ is the derivative of $f$.  Since $f'$ is
classical and $a$ is standard, it follows that $f'(a)$ is standard,
and in particular it is not large.  Therefore, for any $x_1$ such that
$x_1 \approx a$ and $x_1 \ne a$, the difference quotient at $x_1$ must be
close to $f'(a)$ (by definition of derivative). It follows then that
the difference quotient at $x_1$ is not large, since it's close to something
that is not large. Moreover, since $\approx$ is transitive, if $x_2$
is also such that $x_2 \approx a$ and $x_2 \ne a$, then the
difference quotients at $x_1$ and $x_2$ are both close to $f'(a)$, so they
must also be close to each other. Thus, $f$ is differentiable
according to the non-standard criterion.  This simple argument is
sufficient to combine the results of differentiability in ACL2(r) with
the automatic differentiator described
in~\cite{ReGa:automatic-differentiator}, making the automatic
differentiator much more useful, since the notion of differentiability it uses
is now consistent with the main definition in ACL2(r).

Next, we show that the non-standard definition of derivative is
equivalent to the traditional definition (both for the hyperreals and
for the standard reals).  The proof is nearly identical to the
corresponding proof about limits, so we omit it here.

\subsection*{Discussion}

There is a possible misconception that needs to be corrected.  We have
shown that the three different notions of differentiability are
equivalent in principle. However, this is far from sufficient in
practice.

To understand the problem, consider a function such as $x^n$, which
may be represented in ACL2(r) as \texttt{(expt x n)}.  In a real
application of analysis, we may want to show that $f(x) = x-x^{2n}$
achieves its maximum value at $x=1/\sqrt[2n-1]{2n}$.  ACL2(r) has the
basic lemmas that are needed to do this:
\begin{itemize}
  \item $\frac{d(x^n)}{dx} = n \cdot x^{n-1}$ (at least for standard $n$)
  \item Chain rule
  \item Extreme value theorem (EVT)
  \item Mean value theorem (MVT)
\end{itemize}
But these lemmas cannot be used directly.  Consider the chain rule,
for example.  Its conclusion is about the differentiability of $f
\circ g$, and the notion of differentiability is the non-standard
definition. What this means is that the functions $f$ and $g$ cannot
be instantiated with pseudo-lambda expressions, so $f$ and $g$ must be
unary, and that rules out $x^n$ which is formally a binary function,
even if we think of it as unary because $n$ is fixed.

Moreover, suppose that we have a stronger theorem, namely that 
\begin{equation*}
\frac{d(x^n)}{dx} = n \cdot x^{n-1}
\end{equation*}
for all $n$, not just the standard ones.  It's possible to prove this
using induction and the hyperreal definition of differentiability
(since it's a purely classical definition, so induction can be used
over all the naturals, not just the standard ones). Suppose we want to
invoke the MVT on $x^n$ over some interval $[a,b]$. It is not possible
to use the equivalence of the hyperreal and non-standard
definitions. The reason, again, is that the non-standard definition is
non-classical, so we cannot use pseudo-lambdas in functional
instantiations. Even though the two definitions of differentiability
are equivalent for arbitrary (unary) $f(x)$, they are not equivalent 
for the function $x^n$ (which is binary).

It may seem that this is an unnecessary limitation in the part of
ACL2(r). But actually, it's just part of the definition.  The
non-standard definition says that the difference quotient of $f$ is close to
$f'$ at standard points $x$.  It says nothing about non-standard
points.  But when a binary function is considered, e.g., $x^n$, what
should happen when $x$ is standard but $n$ is not?  In general, the
difference quotient need \emph{not} be close to the derivative.

This fact can be seen quite vividly by fixing $x=2$ and $N$ an
arbitrary (for now), large natural number.  Is the derivative
with respect to $x$ of $x^n$ close to the difference quotient when $x=2$ and
$n=N$?   The answer can be no, as the following derivation shows:
\begin{align*}
\frac{(2+\epsilon)^N - 2^N}{\epsilon} &= \frac{(2^N + N \epsilon
  2^{N-1} + {N \choose 2} \epsilon^2 2^{N-2} + \cdots + \epsilon^N) - 2^N}{\epsilon}\\
 & = \frac{N \epsilon  2^{N-1} + {N \choose 2} \epsilon^2 2^{N-2} + \cdots + \epsilon^N}{\epsilon}\\
 & = \frac{\epsilon(N  2^{N-1} + {N \choose 2} \epsilon 2^{N-2} + \cdots + \epsilon^{N-1}}{\epsilon}\\
 & = N  2^{N-1} + {N \choose 2} \epsilon 2^{N-2} + \cdots + \epsilon^{N-1}\\
\end{align*}
All terms except the first have a factor of $\epsilon$, so if $N$ were
limited, those terms would be infinitesimally small, and thus the
derivative would be close to the difference quotient.  But if $N$ is large,
${N \choose 2} = \frac{N(N-1)}{2}$ is also large.  And if $N =
\ceil{1/\epsilon}$, then ${N\choose2}\epsilon$ is roughly $N/2$, which is
large.  So the difference between the difference quotient and the derivative
is arbitrarily large!

This shows that it is not reasonable to expect that we can convert
from the traditional to the non-standard definition of derivative in
all cases. Therefore, we cannot use previously proved results, such as
the MVT directly.

A little subterfuge resolves the practical problem. What must be done
is to prove a new version of the MVT (and other useful theorems about
differentiability) for functions that are differentiable according to
the $\epsilon$-$\delta$ criterion for reals or hyperreals, as
desired. Of course, the proofs follow directly from the earlier
proofs.  For instance, suppose that $f(x)$ is differentiable according to
the hyperreal criterion. Then, we can use the equivalence theorems to
show that $f(x)$ is differentiable according to the non-standard
criterion. In turn, this means that we can prove the MVT for $f(x)$
using functional instantiation.  Now, the MVT is a classical
statement, so we instantiate it functionally with pseudo-lambda
expressions.  E.g., we can now use the MVT on $f(x) \rightarrow
(\lambda (x) x^n)$. So even though we cannot say that $x^n$ satisfies
the non-standard criterion for differentiability, we can still use the
practical results of differentiability, but only after proving
analogues of these theorems (e.g., IVT, MVT, etc.) for the classical versions of
differentiability.  The proof of these theorems is a straightforward
functional instantiation of the original theorems.  We have done this
for the key lemmas about differentiation (e.g., MVT, EVT, Rolle's
Theorem, derivative composition rules, chain rule, derivative of
inverse functions). We have also done this for some of the other
equivalences, e.g., the Intermediate Value Theorem for continuous
functions.

\section{Integrability of Functions}
\label{integrability}

The theory of integration in ACL2(r) was first developed
in~\cite{Kau:ftc}, which describes a proof of a version of the
Fundamental Theorem of Calculus (FTC). The version of the FTC
presented there is sometimes called the First Fundamental Theorem of
Calculus, and it states that if $f$ is integrable, then a function $g$
can be defined as $g(x) = \int_{0}^{x}{f(t) dt}$, and that $g'(x) =
f(x)$. As part of this proof effort, we redid the proof
in~\cite{Kau:ftc}, and generalized the result to what is sometimes
called the Second Fundamental Theorem of Calculus.  This more familiar
form says that if $f'(x)$ is continuous on $[a,b]$, then
$\int_{a}^{b}{f'(x) dx} = f(b) - f(a)$.

The integral formalized in~\cite{Kau:ftc} is the Riemann integral, and
the non-standard version of integrability is as follows:
\begin{equation*}
\int_{a}^{b}{f(x) dx} = L \Leftrightarrow (\forall P)
  \left(P \text{ is a partition of } [a, b] \wedge small(||P||) \Rightarrow \Sigma_{x_i \in P}
    \left(f(x_i) (x_{i} - x_{i-1})\right) \approx L \right)
\end{equation*}
$P$ is a monotonically increasing partition of $[a,b]$ if $P$ is given by a list $P = [ x_1,
x_2, \dots, x_n]$ with $x_1=a$ and $x_n=b$.  The term $||P||$ denotes the maximum value of $x_i -x_{i+1}$
in the partition $P$.

The traditional definition uses limits instead of the notion of
infinitesimally close.  It can be written as follows:
\begin{equation*}
\int_{a}^{b}{f(x) dx} = L \Leftrightarrow
\lim_{||P|| \rightarrow 0}
  \left(\Sigma_{x_i \in P}
    \left(f(x_i) (x_{i} - x_{i-1})\right) \right) \approx L.
\end{equation*}
The notion of limit is strange here, because what approaches 0 is
$||P||$.  Many partitions can have the same value of $||P||$, so this
limit ranges over all such partitions at the same time.

Opening up the definition of limits, integrals can be
expressed as follows: 
\begin{gather*}
\int_{a}^{b}{f(x) dx} = L \Leftrightarrow \\
\qquad (\forall \epsilon>0)(\exists
\delta>0)(\forall P) \\
  \qquad\qquad
  \left(P \text{ is a partition of } [a, b] \wedge ||P|| < \delta \Rightarrow \left|\Sigma_{x_i \in P}
    \left(f(x_i) (x_{i} - x_{i-1})\right) - L\right| < \epsilon\right).
\end{gather*}
Once integrals are viewed in this way, the remainder of the proof is
clear. Specifically, it follows the same line of reasoning as in
Section~\ref{limits}.  First, the $\delta$ that exists depends on $a$,
$b$, and $\epsilon$, so it is standard when those are
standard. Second, since there is a standard $\delta$ that is
sufficient, any infinitesimal can take the place of $\delta$, and then
the condition $||P||<\delta$ can be recast as $small(||P||)$. Finally,
since the Riemann sum is within $\epsilon$ of $L$, for an arbitrary,
positive, standard $\epsilon$, it must be that the Riemann sum is
infinitesimally close to $L$. So the two definitions are, in fact,
equivalent.

\section{Conclusions}
\label{conclusions}

In this paper, we showed how the non-standard definitions of
traditional concepts from analysis are in fact equivalent to the
traditional $\epsilon$-$\delta$ definitions. The results are
especially important in ACL2(r) because the non-standard definitions
feature non-classical notions, such as ``infinitely close'' and
``infinitely small.'' Consequently, they are limited in the use of
induction and functional instantiation. However, the traditional
notions are (by definition) classical, so they are unencumbered by
such limitations.

This presents an interesting dilemma. In our experience, analysis style
proofs are much easier to do and automate using non-standard
analysis. However, \emph{using} those results in subsequent proof
attempts is much easier to do with the traditional (i.e., classical)
statements. The distinction we're making is between \emph{proving} the
correctness of Taylor's Theorem, say, and actually
\emph{using} Taylor's Theorem in a larger verification effort. For
example, the formalization of Taylor's Theorem in~\cite{SaGa:sqrt}
took extreme care to push free variables (including what were really
summation indexes for the series) all the way into the original
\texttt{encapsulate} introducing the function to be
approximated. However, now that the equivalences are proved, a more
elegant approach can be followed: First, prove a ``clean'' version of
Taylor's Theorem using NSA, then use that result to show that Taylor's
Theorem also holds using the traditional definition of derivative. The
``traditional'' version of Taylor's Theorem would then be used with no
restrictions during functional instantiation, so free variables would
no longer present a problem.  We plan to pursue this idea for Taylor's
Theorem in the near future, as part of a comprehensive verification effort 
into the implementation of hardware algorithms for square root
and various trigonometric and exponential functions.

\bibliographystyle{eptcs}
\bibliography{rag}

\end{document}